\documentclass[english, a4paper]{article}

\usepackage[T1]{fontenc}
\usepackage{babel}
\usepackage{amsmath}
\usepackage{amssymb}
\usepackage{graphicx}
\usepackage{epsfig}

\makeatletter
\@addtoreset{chapter}{part}
\makeatother

\newcommand\lsim{\mathrel{\rlap{\lower4pt\hbox{\hskip1pt$\sim$}}
    \raise1pt\hbox{$<$}}}
\newcommand\gsim{\mathrel{\rlap{\lower4pt\hbox{\hskip1pt$\sim$}}
    \raise1pt\hbox{$>$}}}
    
\newcommand\be{\begin{equation}}
\newcommand\bea{\begin{eqnarray} \nonumber }
\newcommand\ee{\end{equation}}
\newcommand\eea{\end{eqnarray}}

\begin{document}

\title{Deconstructing the Low-Vol Anomaly}

\author{S. Ciliberti, Y. Lemp\'eri\`ere, A. Beveratos, G. Simon, \\
L. Laloux, M. Potters, J.-P. Bouchaud \\
Capital Fund Management, \\ 
23 rue de l'Universit\'e, 75007 Paris, France \\
}
\date{\today}

\maketitle

\begin{abstract}
We study several aspects of the so-called low-vol and low-$\beta$ anomalies, some already documented (such as the  universality of the effect over different geographical zones), others 
hitherto not clearly discussed in the literature. Our most significant message is that the low-vol anomaly is the result of {\it two} independent effects. One is the striking negative correlation between past 
realized volatility and dividend yield. Second is the fact that ex-dividend returns themselves are weakly dependent on the volatility level, leading to better risk-adjusted 
returns for low-vol stocks. This effect is further amplified by compounding. We find that the low-vol strategy is not associated to short term reversals, nor does it qualify as a Risk-Premium strategy, since its overall skewness is slightly positive. 
For practical purposes, the strong dividend bias and the resulting correlation with other valuation metrics (such as Earnings to Price or Book to Price) does make the low-vol strategies to some extent redundant, at
least for equities. 
\end{abstract}

\section{Introduction}

The so-called low-volatility (or low-$\beta$) anomaly has been noticed at least as early as 1970 by Fisher Black \cite{Black} -- who failed to convince the Wells Fargo to launch a levered fund that would 
buy low-volatility stocks and sell high volatility ones -- and in 1972 by Robert Haugen -- who equally failed to have his paper \cite{HH} published before his results contradicting the CAPM model were excised \cite{HH2}. 
That low-volatility stocks should perform better than their 
high-volatility counterpart is indeed counter-intuitive and in blatant contradiction with the idea, deeply rooted in economic theory, that risk should be somehow rewarded by some excess return \cite{CAPM,Absurd}. Still, concurrent 
empirical evidence has accumulated since the early seventies, and broadly confirm that this low-volatility ``puzzle'' is a robust, universal stylized fact of stock markets (and, to a lower extent, of bond markets and other asset classes 
as well \cite{Ang,BAB,UBS}). The effect has indeed
been persistent over time, and is documented on a variety of stock markets throughout the world (developed countries or emerging markets alike), see e.g. \cite{Blitz,BakerHaugen,Chen,Japs}. 

Such a striking departure from the efficient market lore begs for an explanation. Several plausible stories can in fact be found in the literature, as reviewed in \cite{Loh,Clemens,BMO}. Some are based on behavioural biases (lottery ticket investing \cite{Barberis,Vorkink,Lotteries1,Japs,Loh}, ``glittering'' stocks attracting the attention of investors \cite{Odean}, or over-optimism of 
analysts for high-vol stocks \cite{Yamada}), others on institutional constraints (high volatility stocks as an alternative to leverage \cite{Black}) 
or incentives (managers' bonuses are in fact options on the performance of invested stocks, and thus more valuable for high volatility stocks \cite{BakerWurgler,BakerHaugen}), others still based on more
`mechanistic' effects, see \cite{Northern,UBS}. 

The low-volatility (``low-vol'' henceforth) anomaly is thus clearly relevant both from a theoretical and practical point of view: it challenges the pillars of modern academic finance, and suggests interesting defensive stock strategies that would have 
significantly outperformed the market in the last 50 years. As such, it has attracted tremendous interest recently, with dozens of papers appearing in the academic and professional literature, see e.g. Refs. \cite{Ang,Clarke,Blitz,Fu,Ang2,BakerWurgler,BakerHaugen,BAB,Japs,Sullivan,NovyMarx,UBS}. Whereas all these papers confirm that the low-vol anomaly is strong and pervasive in stock markets, the origin of the effect is still debated. Novy-Marx, in 
particular, argues that the low-vol anomaly can actually be subsumed in more classic explanatory variables such as value or 
profitability \cite{NovyMarx}. We had ourselves undertaken an in-depth study of the low-vol effect
when the paper of Novy-Marx came out, and our conclusions partly overlap with his. Still, some of our findings appear to be new and, we hope, help shed light on these matters. Our main results are as follows:

\begin{itemize}
\item We do confirm once again the strength and persistence of the low-vol and low-$\beta$ effect on a pool of 9 different countries; in fact we find that the P\&L of the two anomalies are very strongly correlated ($\rho \approx 0.9$)
suggesting that these two anomalies are in fact one and the same. However, since the market neutral low-vol/low-$\beta$ strategy has (by construction) a long dollar bias, it is sensitive to the financing rate. 

\item We find that the low-vol anomaly has nothing to do with short-term (one month) stock reversal -- at variance with some claims in the literature, as it entirely survives lagging the measure of past volatilities by one month or more. The low-vol effect is therefore a persistent, long-term effect.

\item We find that, as expected, low-vol (low-$\beta$) portfolios have strong sector exposures. However, the performance of these strategies remains strong even when sector neutrality is
strictly enforced. The low-vol effect is therefore not a sector effect.

\item We find that a large proportion of the low-vol performance is in fact eked out from dividends, see Fig. \ref{DYplot} below. This is our central result, that follows from the strong negative correlation between volatility and dividend yields which (oddly) does not seem to be clearly documented in the literature (but see \cite{Clemens} where this correlation is implicitly discussed).
However, the low-vol anomaly persists for ex-dividend returns which are found to be roughly independent of the volatility level. 
Therefore risk-adjusted ex-dividend returns are themselves higher for low-vol stocks, which is in itself an ``anomaly''.

\item We find that the skewness of low-vol portfolios is small but systematically positive, suggesting that the low-vol excess returns cannot be identified with a hidden risk-premium \cite{RiskPremia}.

\item The P\&L of the low-vol strategy is $\sim -0.5$ correlated with the Small-Minus-Big (Size) Fama-French factor, $\sim 0.2$ correlated with the High-Minus-Low (Value) factor and $\sim 0.5$ correlated
with the Earning-to-Price factor, which is expected since earnings and dividends are themselves strongly correlated. Once these factors are controlled for, the residual performance of low-vol becomes insignificant -- see 
Fig. \ref{pnlLOWVOLresi} below. 
This result ties with Novy-Marx's observations \cite{NovyMarx}: profitability measures explain to a large degree the low-vol (low-$\beta$) effect.

\item We find that part of the low-vol effect can be explained by compounding, i.e. the mere fact that a stock having plummeted $-20 \%$ must make $+25 \%$ to recoup the losses. Although significant, this 
mechanism is only part of the story.

\item By analyzing the holding of mutual funds, we find that (at least in the U.S.) these mutual funds are indeed systematically over-exposed to high vol/small cap stocks and underexposed to low-vol/large cap stocks, 
in agreement with the leverage constraint and/or bonus incentives stories alluded to above. A similar observation was made in Ref. \cite{Japs} concerning the behaviour of Japanese institutional investors.
\end{itemize}

Our overall conclusion is that, while the low-vol (/low-$\beta$) effect is indeed compelling in equity markets, it is not a real diversifier in a factor driven portfolio that already has exposure to Value type strategies, in particular Earning-to-Price and Dividend-to-Price. Furthermore, the strong observed dividend bias makes us believe that the effect is probably not as convincing in other asset classes such as bonds, 
although see \cite{BAB} and the discussion in \cite{UBS}.

The outline of the paper is as follows. We first present (Sect. 2) our data and methodology, and broadly confirm  previous findings on the low-vol (/low-$\beta$) effect. We then scrutinize several possible biases (sectoral, financing rates, dividends) in Sect. 3. We decompose the low-vol strategy into deciles and correlate it to other factors in Sect. 4. We briefly discuss the compounding effect in Sect. 5, and analyze the holdings of mutual funds in Sect. 6.

\section{Methodology}

\subsection{Construction of the strategies}

We want to test for the presence of a low-vol or low-$\beta$ effect by creating portfolios that are long low-vol stocks and short high-vol stocks, while having zero (or close to zero) correlation with the market mode. The
performance of such portfolios, and its statistical significance, is a test for accepting or rejecting the existence of these effects in the data. We will consider international pools of stocks, defined more precisely 
in Appendix A. We indicate in Table \ref{poolfig} the starting dates of our simulations in every trading zone, as well as the maximum number of stocks in a given pool. 
As we can see, we cover the main industrialised countries, as well as Brazil, Korea and Hong-Kong. Most pools start around 1996, but we used CRSP data for the US zone to go back to 1966; the end date for all pools is July 16th, 2015. We need four years of data to compute a meaningful correlation matrix for our Markowitz portfolio construction, so the P\&L shown below start in 1970. Also, we note that there is a high level of diversification between these pools, which means that the universality of the effect -- already reported in \cite{Blitz,BakerHaugen,Chen} -- is {\it a priori} non-trivial. Indeed, we find that the low-vol and low-$\beta$ strategies are only weakly ($\rho \sim 0.1$) correlated across geographical zones.

\begin{table}
\begin{center}
\begin{tabular}{|c||c|c||c|}
\hline
\hline
Pool & Starting date & $N$ & Risk-Free rate\\
\hline
Russell 3000 & 1998 & 3000 & USD LIBOR 3m\\
S\&P 500 & 1970 & 500 & USD LIBOR 3m\\
Small Caps. & 1970 & 500 & USD LIBOR 3m\\
Australia & 2002 & 200 & AUD LIBOR 3m\\
UK & 2001 & 100 & GBP LIBOR 3m\\
Europe (ex-UK) & 2002 & 600 & EUR LIBOR 3m\\
Japan & 1993 & 500 & JPY LIBOR 3m \\
Korea & 2002 & 200 & KRW KORIBOR 3m\\
Hong-Kong & 2002 & 440 & HKD HIBOR 3m\\
Brazil & 2001 & 70 & BRL implied 3m \\
Canada & 2001 & 270 & CAD LIBOR 3m\\
\hline
\hline
\end{tabular}
\caption{\small{Starting date of the P\&L simulations 
and number of stocks $N$ for every pool used in the simulation. In the last column we give the financing rates used in the simulations $r_{\text{RF}}$. For Australia and Canada, LIBOR is replaced 
by Accepted Bank Bills since 2013.}}
\label{poolfig}
\end{center}
\end{table}

We first need to measure the volatility ($\sigma_i$) and the ``beta'' ($\beta_i$) of each stock. The former quantity is obtained as a 100-day rolling standard deviation of the stocks total return. The $\beta$s are a bit less 
immediate to measure. We first define, for simplicity, an ``index'' return as an equi-weighted average of all the pool's members (other definitions would anyway lead to a highly correlated strategy, so our definition is just convenient). We then compute $\beta$ as the covariance of the stock with the index divided by the index variance, both computed over 100 days as well, but considering 3-days returns for the variances and covariances as to take into account any lead-lag effects. We further {\it lag these values by an extra one month} (20 trading days). Our volatility indicator is therefore not sensitive to recent returns at all. This in particular excludes any interpretation of the
low-vol anomaly reported below in terms of short-term reversal of strong recent rallies, as proposed in several papers, e.g. \cite{Fu,Sullivan}.\footnote{We therefore strongly disagree with the following statement in \cite{Sullivan}:
{\it In particular, we find that the excess return associated with forming the low risk zero-cost portfolios are short-lived as they are
present only in month $t+1$ and furthermore are largely subsumed by high transaction
costs.}, as also claimed in e.g. \cite{Fu}.}

We now compute the signals $s_i$ based on these quantities. To control the cross-sectional distribution of our predictions over time, as well as to limit the amount of noise, we use the rank of the volatility or $\beta$, rescaled between -1 and 1. In more mathematical terms: 
\be
s_i = \frac2N \,\text{rank}\,\left(\frac{1}{\sigma_i}\right) - 1
\ee
where $s_i$ is the signal on stock $i=1,\dots, N$, $\sigma_i$ its volatility (as defined above), and $N$ the number of stocks in the pool. We take the inverse of the volatility since we want to be long on \emph{low} vol stocks. 
We apply a similar procedure for the $\beta$ strategy. 

However, a portfolio constructed blindly using this signal would end up having a net short market bias, since by construction the long leg of the portfolio is less volatile and has less market exposure than the short leg. To compensate for that effect, we project out the market mode (using the empirical correlation matrix, again computed using past data) which enforces market neutrality. 
An interesting consequence of our portfolio construction is that the final dollar positions do not add up to zero, although the initial signal $s_i$ did. 
Instead, the resulting net dollar exposition is positive -- see Appendix B. In words, this is because of the re-leveraging of the (less volatile) longs that is needed to ensure the market neutrality. But this net long dollar bias must be financed in order to get meaningful P\&L's. We therefore need a history of risk-free rate $r_{\text{RF}}$, which we have listed in the last column of Table \ref{poolfig}. It is mostly the LIBOR 3-months when available, but in Brazil for example, we had to rely on implied rates. We have quite long histories for these rates, so that we are not limited when we perform our back-tests. Therefore, we are able to simulate the effects we want to test over a wide variety of pools. 

The final P\&L of the low-vol/low-$\beta$ strategies is obtained by summing over time and over stocks the (market neutral) positions for each stock, times the return of that stock minus the risk-free rate $r_{\text{RF}}$. 
The resulting time series is what we will call the ``performance'' of the fully-financed strategy. Note that we do not account for transaction costs (that depend on many extra assumptions, in particular on the size of the portfolio), 
difference of financing costs between the long and the short legs of the portfolio, and dividend taxes (although we briefly discuss their impact below). These effects are usually not considered in the academic literature either, 
although see \cite{Sullivan}.

\subsection{Analyzing the low-vol/low-$\beta$ performance}

We now present the result of the simulations we have run, and discuss their significance and robustness. The results are summarized in Table \ref{SRvol}, both for the low-vol and low-$\beta$ anomalies. 
As already documented in \cite{Blitz,BakerHaugen,Chen}, the performance of low-vol/low-$\beta$ strategies appears to very robust across all zones. In fact, there is little dispersion in the performance per pool. 
In particular, {\it all} pools return a positive contribution for both strategies, with no particular bias towards developed or emerging countries. 

\begin{figure}
\begin{center}
\epsfig{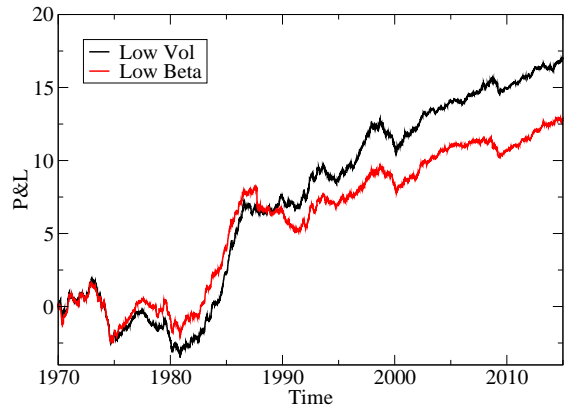}
\caption{\small{Performance of the low-vol and low-$\beta$ portfolios, aggregated across all zones with equal weight on all pools alive at any given date. The period 1970-1990 is US only and is also
a period of high interest rates until 1980.}}
\label{pnlBAV}
\end{center}
\end{figure}

When aggregated in a global portfolio, these strategies yield a total performance plotted in Fig.~\ref{pnlBAV}. Even if for each individual market the statistical significance of low-vol and low-$\beta$ is 
not impressive, the global portfolios have Sharpe ratios of 0.86 and 0.74 respectively (see Table \ref{SRvol}), 
which give them a high statistical significance (t-stat) of 5.8 and 5.0.  The performance of low-vol/low-$\beta$ is furthermore only weakly correlated $(\sim 0.1)$ between different geographical zones, and the 
mutual correlation between  low-vol/low-$\beta$ is as high as 0.88. This leads us to conclude that these two anomalies are in fact very closely related, and 
we will not really distinguish them in the following discussion. A simple argument explaining why low-vol/low-$\beta$ are so strongly correlated is given in Appendix C. 
The reader must also have noticed that we consider the total volatility of each stock, including that of the market. Some authors prefer to focus on the residual contribution only when defining low-vol and high-vol
stocks. However, this also leads to a strategy that is highly correlated to our definition of low-vol and to low-$\beta$; see again Appendix C for why this is the case.

We have also tested different portfolio constructions, and always found results in line with the above. Therefore, we strongly believe that the results presented here are not due to an artifact of our specific 
methodology. 

Finally, we have computed the skewness of the performance for all zones. In order to reduce measurement noise, we have chosen to compute the skewness as the mean minus the median of daily returns, divided by 
the rms of the returns. We find a small, but systematically positive skewness for all zones. Following our discussion in Ref. \cite{RiskPremia}, this suggests that the excess return associated to low-vol 
stocks {\it cannot} be interpreted as a hidden risk premium, and is probably of behavioural origin, see e.g. \cite{Loh, Clemens} and the discussion below.

\begin{table}[htbp]
 \resizebox{0.92\textwidth}{!}{\begin{minipage}{\textwidth}
\begin{tabular}{|c||c|c|c||c|c||c|}
\hline
\hline
Pool & low-vol & low-$\beta$  & low-vol-SN & corr 1-2 & corr. 1-3 & skewness \\
\hline
Russell 3000 & 0.67 & 0.80 & 0.53 & 0.86 & 0.88 & 0.0 \\
S\&P 500 & 0.24 & 0.14 &  0.13 & 0.72 & 0.77 & 0.02 \\
Small Caps. & 0.87 & 0.94 & 0.74 & 0.86 & 0.77 & 0.0 \\
Australia & 0.64 & 0.70 & 0.48 & 0.71 & 0.78 & 0.03 \\
UK & 0.35 & -0.08 & -0.07 & 0.66 & 0.67 & -0.01 \\
Europe (ex UK) & 0.76 & 0.50 & 0.71 & 0.79 & 0.88 & 0.01 \\
Japan & 0.45 & 0.48 & 0.24 & 0.75 & 0.82 & 0.02 \\
Korea & 0.86 & 0.56  & 0.77 & 0.66 & 0.85 & 0.02 \\
Hong-Kong & 1.00 & 1.01  & 0.74 & 0.78 & 0.91 & 0.01 \\
Brazil & 0.45 & 0.39 & 0.02 & 0.73 & 0.72 & 0.02 \\
Canada & 0.71 & 0.73 & 0.51 & 0.77 & 0.74 & 0.0 \\
\hline
\hline
Average & 0.63  & 0.56   &  0.43 & 0.83 & 0.80 & 0.01  \\
\hline
\hline
Global & 0.86 & 0.74 & 0.62 & 0.83 & 0.82 & 0.03 \\
\hline
\hline
\end{tabular}
\end{minipage}}
\caption{\small{Sharpe ratio for low-vol, low-$\beta$ anomalies and sector-neutral (SN) low-vol. Although not very significant at the single index level (t-stats around $1.5 - 2$), the performance is robust 
and consistent across geographical zones, nor very much relying on sectoral biases. We also give the correlations between low-vol and low-$\beta$ (corr 1-2), and between low-vol and low-vol-SN (corr 1-3). 
Finally, we give the skewness of the low-vol strategy in the last column. 
The last line refers to the results of the strategy applied with equal weight to all pools alive at any given date, also shown in Fig.~\ref{pnlBAV}, with a t-stat above $5$. 
Note that the statistics is dominated by the US market until 2000. The Sharpe ratio of the global low-vol strategy after 2002 actually rises to $1.41$ (t-stat $=4.9$).}}
\label{SRvol}
\end{table}

\section{Three possible biases}

\subsection{Dollar bias and financing costs}

As mentioned above, the re-leveraging of the long leg of the portfolio needed to ensure the market neutrality leads to a net long bias. This is actually not totally intuitive
and a simple illustrative model that accounts for this effect is provided in Appendix B. The financing cost 
was accounted for in the simulation results provided above by systematically subtracting $r_{\text{RF}}$ from the stock returns. 
However, the difference between lending and borrowing rates was neglected in our 
study, as in most other academic studies. 

We found no correlation between the financing rate level, and the un-financed performance of the low-vol (/low-$\beta$) strategy. Therefore, it seems plausible that because of the net-long bias, 
these strategies would stop working if the risk-free rate became too high. Obviously, in the 2015 situation, we are far from this case, but it could help understand the relatively poor performance that 
is observed in the US in the 70-80s. (Note that the P\&L displayed in Fig.~\ref{pnlBAV} is entirely coming from the US stocks until 1998).

\subsection{Sector biases}

There is considerable evidence (as well as strong intuitive reasons to believe) that the anomalies we study have persistent sectoral biases. In particular, we expect our low-vol portfolios to be long Utilities/Consumer Non-Cyclical, and short Technologies/Consumer Cyclical. We checked that this is indeed the case on all our zones. 

Now, an interesting question arises: are these anomalies merely exploiting sector biases? Are they simply picking up the sectors with the best risk-adjusted return? We decided to confront this issue directly
by building sector neutral portfolios, i.e. we rank the volatility over every sector between -1 and 1, and then apply our usual portfolio construction. By definition, these portfolios should not -- and indeed do not -- 
have any sectoral bias left. 

We summarize the performance of this strategy for low-volatility in Table~\ref{SRvol} (results for low-$\beta$ would lead to similar conclusions). 
As we can see, the Sharpe ratios are only slightly reduced, and the correlation is still very high between the two implementations. This means that the sector component is not 
a dominant determinant of the low-vol/low-$\beta$ anomaly. 

\subsection{Dividend bias}

\begin{figure}
\begin{center}
\epsfig{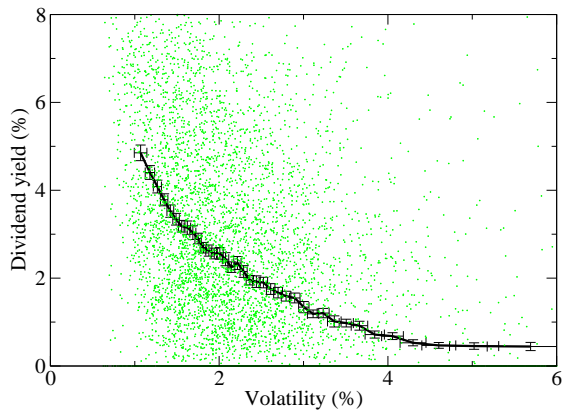}
\caption{\small{Scatter plot of the dividend yield (DY, in \%) of the 2000 largest US companies since 1971 as a function of the past 250-day volatility $\sigma$ (in \%).
(one point per year and per stock, but only one random point out of ten shown on the graph, for clarity).  
We find a clear negative correlation around $-0.2$ between dividends and volatility, which is illustrated by the binned averages shown as the black line. 
Error bars are computed for each bin, which contain 2000 points. Earning-to-Price ratio of SPX companies as a function of the past volatility show a very similar pattern.}}
\label{DYplot}
\end{center}
\end{figure}

A rather striking observation (that, curiously, we have not seen clearly stated in the literature before) is that most of the gain of the low-vol strategy in fact comes 
from the dividend part, as can be seen from Table~\ref{DYratio}. Since the portfolio is market neutral, it must be that its long, low-vol stocks receive on average higher dividends than short, high-vol stocks. 
Indeed, we have checked that there is a significant negative causal correlation ($\sim -0.2$ for US stocks) between past realized volatility and dividend yields, see Fig.~\ref{DYplot}. 
This demonstrates that low dividend stocks are also on average high dividend yielders. 

Why is this so? One argument could be that since high-vol, ``glittering''
stocks are attractive because of all the biases mentioned in the introduction, low-vol ``boring'' stocks must somehow compensate by offering larger dividends. 
A slight variation on this idea is that mature businesses pay dividends whereas growing (risky) firms do not.  
However, the causality could be reversed: not surprisingly, we find a similar negative correlation between earnings and volatility. One could thus
argue that strong earnings and regular dividends make firms less risky and therefore less volatile. 

\begin{table}
\begin{center}
\begin{tabular}{|c|c|}
\hline
\hline
Pool & dvd/total gain \\
\hline
Russell 3000 & 88 \% \\
S\&P 500 & 261 \% \\
Small Caps. & 125 \% \\
Australia & 75 \% \\
UK & 64 \% \\
Europe (ex-UK) & 44 \% \\
Japan & 24 \% \\
Korea & 17 \%\\
Hong-Kong & 51 \%\\
Brazil & 50 \% \\
Canada & 59 \% \\
\hline
\hline
\end{tabular}
\caption{\small{Contribution of the dividend gains (dvd) to the performance of low-vol strategies (total gain). Very similar figures are obtained for low-$\beta$ as well.}}
\label{DYratio}
\end{center}
\end{table}

In any case, this observation leads to the concern that part of the low-vol or low-$\beta$ performance might eventually be eaten up by dividend taxes. The tax rate is however very much investor-dependent; our analysis suggests that the low-vol strategy can in fact withstand moderate dividend tax levels (up to $50 \%$) before becoming flat. 
Nonetheless, this is another concern for the viability of these strategies, on top of the level of interest rates to finance the leveraged positions. 

\begin{figure}
\begin{center}
\epsfig{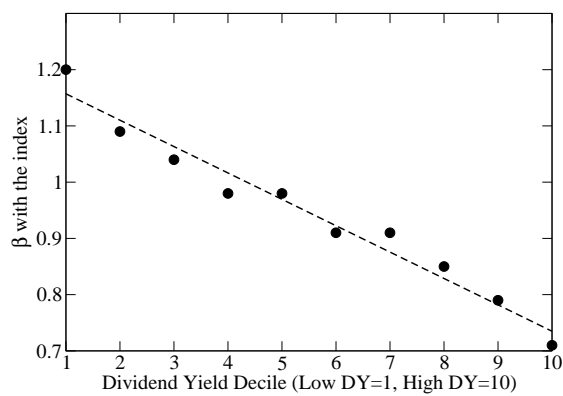}
\caption{\small{$\beta$s of the 10 Fama-French Dividend Yield portfolios since 1950. The dashed line is a linear regression through the 10 points, as a guide to the eye. 
As one can see, the higher the yield, the smaller the $\beta$ with the index, as expected from the 
negative volatility/dividend  correlation.}}
\label{FFDYdec}
\end{center}
\end{figure}

Incidentally, the above results suggest that simple dividend yield strategies need to be carefully risk-controlled if one wants to avoid any market exposure. Indeed, since high dividend yield stocks are also low-vol/$\beta$, one would expect to end up with a short market exposure if longs are not re-leveraged. This is exactly what happens to the Fama-French portfolios based on the Dividend Yield (DY) factor, see  Fig.~\ref{FFDYdec}, where we have plotted the $\beta$ of each of the 10 
Fama-French decile portfolios created since 1950.  As one can see, the highest ranking ones have a much smaller $\beta$ compared to the others. 
This also means that going equally long the high dividend portfolios, and short the low ones results in a significantly negative correlation with the index of around $-0.4$.

\section{The low-vol strategy: Deciles and Factors}

\subsection{Deciles} 

\begin{figure}
\begin{center}
\epsfig{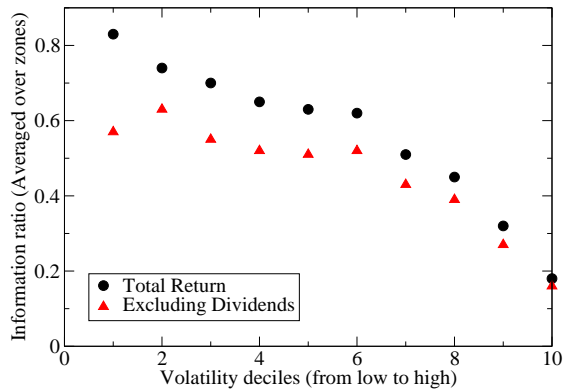}
\caption{\small{Average over all zones of the Information Ratios of the total return (black circles) and ex-dividend returns (red triangles) for the deciles portfolios, now from 
low-vol to high-vol. The average is computed as a flat average over the columns of Tables \ref{Decile-total}, \ref{Decile-exdiv}. One clearly sees that risk adjusted, ex-dividend returns 
are slightly better for low-vol stocks than for high-vol stocks. The effect is amplified by dividends.}}
\label{Fig_deciles}
\end{center}
\end{figure}

In the following Tables, we give the Information Ratio (i.e. the Sharpe ratio of the un financed strategy) of Fama-French like past volatility ``decile" portfolios, for all geographical zones, for both the total return strategy
and for the ex-dividend strategy.\footnote{We use the same definition of the volatility as above to rank the stocks: we mean the past 100 days realized volatility, lagged by one month.} One sees that, as expected from positive performance of the low-vol strategy, the Information Ratio of low-vol portfolios is significantly higher than that of high-vol portfolios, 
{\it even when dividends are left out} -- see Table \ref{Decile-exdiv}; in particular the last line that gives the average over all zones and Fig. \ref{Fig_deciles}. This means that there is a genuine low-vol effect here, 
on top of the strong dividend effect noted in the previous paragraph. We see that the anomaly is not localized on any of the ten deciles, but is rather a smooth bias that builds up progressively as one moves from high-vol stocks to low-vol stocks. This remark is at odds with claims in the literature that the anomaly is chiefly due to penny stocks, or extremely volatile stocks that plummet.

\begin{table}[htbp]
\resizebox{0.88\textwidth}{!}{\begin{minipage}{\textwidth}
\begin{tabular}{|c|c|c|c|c|c|c|c|c|c|c|}
\hline
\hline
Pool & 1 & 2 & 3 & 4 & 5 & 6 & 7 & 8 & 9 & 10 \\
\hline
Russell 3000 & 0.25 & 0.42 & 0.50 & 0.54 & 0.57 & 0.60 & 0.64 & 0.66 & 0.70 & 0.84 \\
S\&P 500     & 0.49 & 0.59 & 0.71 & 0.73 & 0.79 & 0.78 & 0.88 & 0.86 & 0.90 & 1.01 \\
Small Caps.  & 0.39 & 0.54 & 0.71 & 0.72 & 0.87 & 0.90 & 1.12 & 1.05 & 1.15 & 1.34 \\
Australia    & 0.18 & 0.18 & 0.51 & 0.51 & 0.65 & 0.69 & 0.73 & 0.79 & 0.88 & 0.90 \\
UK           & 0.13 & 0.26 & 0.21 & 0.29 & 0.56 & 0.52 & 0.27 & 0.47 & 0.94 & 0.64 \\
Europe (ex-UK)      & 0.28 & 0.44 & 0.53 & 0.55 & 0.59 & 0.60 & 0.62 & 0.67 & 0.74 & 0.91 \\
Japan       & -0.16 & 0.16 & 0.24 & 0.23 & 0.26 & 0.25 & 0.35 & 0.30 & 0.33 & 0.30 \\
Korea       & -0.35 & -0.10 & 0.28 & 0.55 & 0.67 & 0.73 & 0.57 & 0.75 & 0.65 & 0.67 \\
Hong-Kong    & 0.49 & 0.67 & 0.75 & 0.66 & 0.71 & 0.70 & 0.73 & 0.98 & 1.03 & 1.15 \\
Brazil       & 0.22 & 0.25 & 0.23 & 0.41 & 0.49 & 0.47 & 0.43 & 0.41 & 0.76 & 0.59 \\
Canada       & 0.09 & 0.06 & 0.29 & 0.42 & 0.61 & 0.66 & 0.80 & 0.77 & 0.95 & 0.81 \\
\hline
\hline
\end{tabular}
\end{minipage}}
\caption{\small{Information Ratios of the ``total return'' performance (i.e. including dividends)  of past volatility decile portfolios. (The first decile is the most volatile)}}
\label{Decile-total}

\end{table}

\begin{table}[htbp]
 \resizebox{0.88\textwidth}{!}{\begin{minipage}{\textwidth}
\begin{tabular}{|c|c|c|c|c|c|c|c|c|c|c|}
\hline
\hline
Pool & 1 & 2 & 3 & 4 & 5 & 6 & 7 & 8 & 9 & 10 \\
\hline
Russell 3000 & 0.25 & 0.40 & 0.48 & 0.51 & 0.53 & 0.54 & 0.56 & 0.56 & 0.56 & 0.59 \\
S\&P 500     & 0.45 & 0.53 & 0.61 & 0.61 & 0.65 & 0.61 & 0.70 & 0.65 & 0.65 & 0.63 \\
Small Caps.  & 0.38 & 0.53 & 0.68 & 0.67 & 0.80 & 0.80 & 0.97 & 0.84 & 0.84 & 0.84 \\
Australia    & 0.12 & 0.10 & 0.36 & 0.35 & 0.44 & 0.45 & 0.49 & 0.51 & 0.55 & 0.54 \\
UK           & 0.08 & 0.17 & 0.11 & 0.17 & 0.42 & 0.36 & 0.12 & 0.24 & 0.72 & 0.39 \\
Europe (ex-UK)       & 0.26 & 0.39 & 0.45 & 0.46 & 0.48 & 0.48 & 0.47 & 0.53 & 0.57 & 0.71 \\
Japan       & -0.18 & 0.13 & 0.20 & 0.18 & 0.21 & 0.19 & 0.28 & 0.23 & 0.25 & 0.21 \\
Korea       & -0.36 & -0.11 & 0.26 & 0.52 & 0.62 & 0.69 & 0.52 & 0.70 & 0.58 & 0.58 \\
Hong-Kong    & 0.44 & 0.61 & 0.68 & 0.58 & 0.61 & 0.58 & 0.60 & 0.84 & 0.83 & 0.83 \\
Brazil       & 0.20 & 0.22 & 0.18 & 0.36 & 0.42 & 0.39 & 0.35 & 0.31 & 0.62 & 0.43 \\
Canada       & 0.08 & 0.04 & 0.24 & 0.33 & 0.49 & 0.51 & 0.65 & 0.59 & 0.73 & 0.57 \\
\hline
\hline
\end{tabular}
\end{minipage}}
\caption{\small{Information Ratios of the price return (ex-dvd) performance of past volatility decile portfolios. (The first decile is the most volatile).}}
\label{Decile-exdiv}
\end{table}

\subsection{Skewness}

We have also reported in Table \ref{Decile-skew} the skewness of the different decile portfolios. 
One notices that all skewnesses are negative but that the skewness of high-vol portfolios is slightly less negative than that of the low-vol ones. 
This is somewhat unexpected since, as we have noted before,
buying low-vol stocks and shorting high-vol stocks leads to a {\it positively} skewed strategy (see last column of Table \ref{SRvol}). This is of course possible because skewnesses do not simply add when 
returns are correlated. Following up on this, 
we have checked on all geographical zones that while the average return of high-vol stocks is better than that of low-stocks on days where the index goes up, a
stronger opposite effect is observed on days when the index goes down. More precisely the (negative) return differential between high-vol and low-vol stocks is $\approx 1.5$ 
times larger when the index goes down -- leading to strong positive gains that contribute to the positive overall skewness of the low-vol strategy. This is why  
these strategies are often called ``defensive'': they allow to make profits in bear market environments. 

\begin{table}[htbp]
 \resizebox{0.83\textwidth}{!}{\begin{minipage}{\textwidth}
\begin{tabular}{|c|c|c|c|c|c|c|c|c|c|c|}
\hline
\hline
Pool & 1 & 2 & 3 & 4 & 5 & 6 & 7 & 8 & 9 & 10 \\
\hline
Russell 3000 & -0.59 & -0.49 & -0.29 & -0.28 & -0.36 & -0.41 & -0.38 & -0.39 & -0.48 & -0.39 \\
S\&P 500 & -0.10 & -0.24 & -0.28 & -0.33 & -0.26 & -0.17 & -0.15 & -0.22 & -0.16 & -0.07 \\
Small Caps. & -0.20 & -0.19 & -0.32 & -0.34 & -0.49 & -0.44 & -0.48 & -0.41 & -0.54 & -0.28 \\
Australia & -0.11 & -0.61 & -0.56 & -0.57 & -0.44 & -0.38 & -0.58 & -0.43 & -0.68 & -0.43 \\
UK & -0.16 & -0.13 & -0.41 & 0.33 & -0.24 & -0.29 & -0.22 & -0.63 & -0.20 & -0.58 \\
Europe (ex-UK)  & -0.71 & -0.55 & -0.51 & -0.36 & -0.36 & -0.41 & -0.38 & -0.39 & -0.51 & -0.32 \\
Japan & 0.03 & -0.15 & -0.04 & -0.03 & -0.22 & -0.06 & -0.09 & -0.25 & -0.07 & 0.10 \\
Korea & -0.76 & -0.65 & -0.61 & -0.70 & -0.52 & -0.48 & -0.51 & -0.58 & -0.31 & -0.63 \\
Hong-Kong & -0.11 & 0.01 & -0.20 & -0.37 & -0.01 & -0.33 & -0.45 & -0.29 & -0.38 & -0.56 \\
Brazil & 0.11 & -0.03 & -0.01 & -0.07 & -0.22 & -0.08 & -0.24 & -0.16 & -0.19 & -0.33 \\
Canada & -0.37 & -0.31 & -0.50 & -0.77 & -0.68 & -0.70 & -0.71 & -0.90 & -0.72 & -0.67 \\
\hline
\hline
Average & -0.27  &  -0.30  &  -0.34   &  -0.32   &  -0.35   &  -0.34   &  -0.38  &  -0.42   &  -0.38  &  -0.38  \\
\hline
\hline
\end{tabular}
\end{minipage}}
\caption{\small{Skewnesses of the total return performance of volatility decile portfolios. (The first decile is the most volatile).}}
\label{Decile-skew}
\end{table}

\subsection{Correlating with standard factors}

It is now time to correlate the performance of low-vol strategies with other, more classical stock strategies. We include in these the usual Fama-French factors 
(which we have recoded using our own portfolio construction methodology): UMD (Momentum), SMB (Size) and HML (Value). 
We also included MKT (the market), since we eventually want to remove any residual correlation with the market in our analysis. Given the results of the previous section on the role of dividends, 
we thought it interesting to study also the correlation with other Value/Valuation type metrics, so we consider Earning-to-Price (E/P) and Dividend-to-Price (D/P) on top of Book-to-Price (i.e. HML). 

We show in Table~\ref{STRATcorr} the correlation of low-vol and low-$\beta$ with all these strategies, computed using the corresponding global P\&Ls (i.e aggregating
all geographical zones) and using monthly data. We find, as expected, a strong anti-correlation with SMB (small cap stocks often have high volatilities), moderate correlation with 
HML and strong positive correlation with either E/P or D/P, again expected from the dividend bias documented above. Not surprisingly, we find no correlation at all or low-vol with UMD, whereas other 
``value'' type strategies such as HML are well known to be anti-correlated with UMD.

\begin{table}
\begin{center}
\begin{tabular}{|c|c|c|c|c|c|}
\hline
  & HML & SMB & UMD & E/P & D/P \\
\hline
low-vol  & 0.21 & -0.56 & 0.01 & 0.51 & 0.63 \\
low-$\beta$ & 0.32 & -0.31 & -0.06 & 0.42 & 0.64 \\
\hline
\end{tabular}
\caption{\small{Correlation of low-vol and low-$\beta$ monthly performance with other significant market anomalies. 
Note that the low-vol and low-$\beta$ strategies are only weakly ($< 0.1$) correlated with the market mode (MKT).}}
\label{STRATcorr}
\end{center}
\end{table}

\begin{figure}
\begin{center}
\epsfig{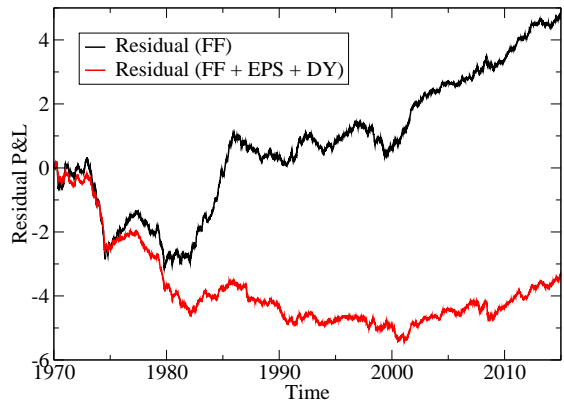}
\caption{\small{Black: Residual of low-vol once the four standard Fama-French factors (MKT, UMD, SMB and HML) are taken into account. The Sharpe ratio of the
residual performance is $\approx 0.4$, but is substantially better ($\approx 0.8$) since the beginning of the 80's when financing rates went down. Red: Residual of low-vol 
once the above four standard Fama-French factors and the E/P and D/P factors are taken into account. The performance becomes flat, at best, since 1980. }}
\label{pnlLOWVOLresi}
\end{center}
\end{figure}

We are now in position to make a residual analysis, extracting the performance of low-vol/low-$\beta$ that is not explained by the above four Fama-French factors. The resulting residual performance 
is plotted in Fig.\ref{pnlLOWVOLresi} for the low-vol strategy (the same results also hold for low-$\beta$). As we can see, though some of the ``alpha'' of the strategies is explained by traditional factors, 
it seems that there is still some additional performance, especially in the recent decades where financing rates were lower. 

However, if we include E/P and D/P as extra factors, the resulting residual P\&L becomes essentially flat over the 1980-2015 period,  
in agreement with Novy-Marx's recent results \cite{NovyMarx}. This residual performance is in fact even quite negative in the 70s, leading to an overall Sharpe ratio on the whole period of around $-0.5$ -- 
see Fig.\ref{pnlLOWVOLresi}.
The operational conclusion is that low-vol is another version of more standard valuation strategies, that can be used as a diversifier in a quant type portfolio, but is not expected to add much ``alpha''.  

\section{A compounding effect?}

A simple idea that could account for the low-vol anomaly is the usual compounding effect \cite{Northern}, is the fact that the geometrical mean is always smaller than 
the arithmetic mean; in more mundane words that $-20 \%$ followed by $+20 \%$ results in a drop of $-4 \%$. This implies that even if the average daily returns of all stocks (high-vol or low-vol)  were exactly equal, the monthly or yearly average return of high-vol stocks would dip below that of low-vol stocks. Low-vol stocks
might be ``defensive'' just because they avoid large drops from which it is hard to recover. This trivial mechanism could be a pervasive reason why one should expect volatile stocks (or any asset for that matter) to under-perform in the long run. 

\begin{figure}
\begin{center}
\epsfig{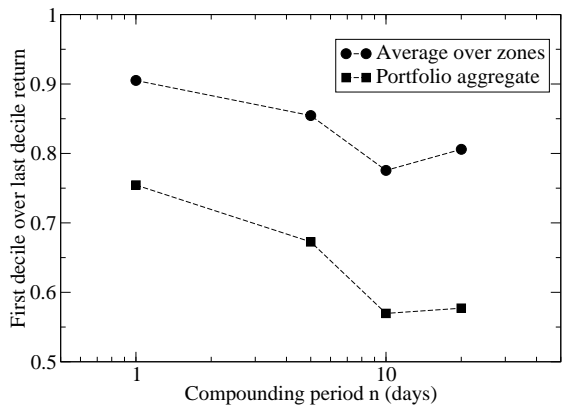}
\caption{\small{First decile (high vol) stocks average price ex-dividend returns, divided by the corresponding last decile (defensive) average price return, over various time-scales $n=1, 5, 10, 20$ days, 
averaged over all zones, or as an aggregate portfolio. In both cases we see that the compounding effect is detrimental to high-vol stocks.}}
\label{decLV}
\end{center}
\end{figure}

In order to assess the relevance of this idea, we have split our portfolio of stocks every day in 10 buckets of decreasing volatility (measured, as above, as 
a 100-days flat average). We then compute the average return (excluding dividends) for each of these deciles over $n$ days, $n=1, 5, 10, 20$. We then plot as a function of $n$ in Fig.~\ref{decLV} 
the ratio of the average return of the first decile (high vol.) to the average return of the last one (low-vol.). As can be seen, ex-dividends daily returns are on average roughly the same for both deciles: 
depending on the geographical zone/market, the ratio is well scattered around unity with an average of $0.9$. As the time
scale over which the return is computed increases, high-vol stocks are seen to perform less and less compared to their defensive counterparts. 
This is consistent with the compounding effect we outlined in the previous paragraphs. 

However, the effect is not strong enough to explain on its own the performance of the low-vol anomaly. Even without the dividend bias, 
the mere fact that all deciles have roughly the same average return at one-day is already at odds with the CAPM, since one would expect high vol stocks to compensate for
risk. Within the CAPM, all stocks should have similar {\it risk-adjusted} returns, at odds with empirical observation -- see again Table \ref{Decile-exdiv} and Fig. \ref{Fig_deciles} above. 

In conclusion, it seems that while the compounding effect does indeed play a part in the low-vol anomaly, it is only part of the explanation: quite apart from
the dividend bias, risk-adjusted returns of highly volatile stocks are already anomalously low.

\section{FactSet data and institutional fund biases}

We now turn to another source of data in search of a plausible explanation of the low-vol anomaly, namely the ``FactSet Equity Ownership'' database
which summarizes the assets of ``institutional, mutual fund, stakeholder, and float-related share ownership information for equities'' (i.e. large un-leveraged institutional players) 
in some of the geographical zones studied in this paper (US, Japan, UK, Europe ex-UK, Australia).  We have aggregated these assets per company name, and normalize this total amount held by mutual funds either by the market capitalization of the company, so as to gauge their overall detention rate of a given name, or by the total holdings of each fund. Both normalisations lead to a similar conclusion, so we focus on the first one. The funds in our data hold on average about 36\% of each company in the SPX, and their total holdings have a sizable standard deviation of about 
10\% around this value. This fraction is found to be 6 \% for the TOPIX in Japan, 23 \% for the FTSE 100 in the UK and our synthetic index in Europe ex-UK, and 10 \% for the S\&P/ASX 200 in Australia. 

\begin{figure}
\begin{center}
\epsfig{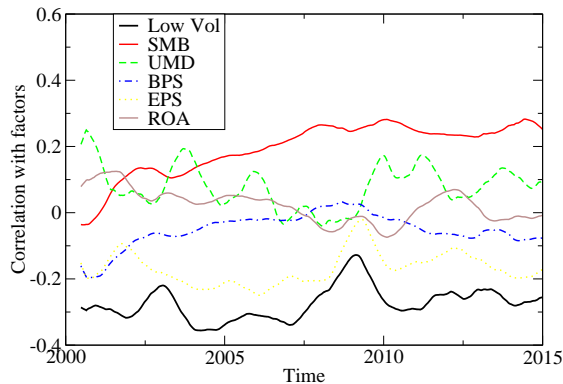}
\caption{\small{1 year running average of the correlation of mutual funds positions with the standard factors: Low Vol, SMB, UMD, BPS, EPS and ROA. We clearly see that these funds are over-allocated 
on small, high vol stocks.}}
\label{avgFS}
\end{center}
\end{figure}

We sample the portfolios of mutual funds every 6 months, and investigate their possible biases by correlating their aggregated positions with various classical factors predictors. The results 
for the US is reported in Fig.~\ref{avgFS}, and other zones behave very similarly. The biases on many standard factors are not clear, but two of them stand out: institutional funds are on average short low-vol and long SMB, i.e. they are over-weighted on high-vol, small cap stocks. 
Although this analysis is obviously not in itself enough to explain why high-vol stocks are overpriced, our results seem consistent with the behavioural/institutional biases alluded to above, i.e. 
large market players/institutions seek high volatility stocks for the embedded leverage they provide, or for the increased probability for the manager to get a better bonus (seen as a kind of European call option on the 
performance of the fund), or else the chronic optimism of analysts that is stronger for high-vol firms than for low-vol firms. 

Similar results hold for all geographical zones, and have been indeed reported for Japanese institutional investors as well in Ref. \cite{Japs}. The SMB bias is even stronger with our second normalisation, 
i.e. when we divide by the total holdings of each fund. This is expected since small funds can afford more weight on 
low cap, illiquid stocks. 

\section{Conclusion}

In this paper, we have dissected several aspects of the so-called low-vol and low-$\beta$ strategies. We have established several ``stylized'' facts about these strategies, some
already well documented (such as the strength and universality of the effect over different geographical zones), others not clearly discussed in the literature before (such as the strong
dividend bias towards low-vol stocks). 

Our most significant message is that the low-vol anomaly is the result of two quite independent structural effects, but of similar strength. One is the strong correlation between low-volatility 
and high dividend yields, so that dividends in fact contribute to a substantial part of the performance of the low-vol strategy. Second is the fact that ex-dividend returns themselves 
are to a first approximation independent of the volatility level on a daily time scale, leading to better risk-adjusted returns for low-vol stocks. This effect is further amplified by
compounding. 

We have also shown that the low-vol anomaly is not localized on extreme volatility deciles and does not come from sectoral biases. It is not due to short-term reversals either, since the volatility 
estimate can be lagged by one month or more without degrading the strength of the anomaly. We furthermore find that the low-vol strategy has a slightly positive overall skewness, 
at variance with standard Risk-Premia strategies that are characterized by negative skewnesses \cite{RiskPremia}. It would actually be very hard to explain intuitively why investing on low-volatility,
high dividend stocks is carrying a specific risk factor that must be compensated for!

For practical purposes, the strong dividend bias and the resulting correlation with other valuation metrics such as E/P or HML does make the low-vol strategies to some extent redundant, at
least for stocks. This does not mean that such strategies should be excluded from the construction of ``quant-factors'' portfolios, since they offer alternative implementations that 
increase diversification and reduce operational risk. 

Finally, the underlying reasons for the low-vol anomaly to persist in equity markets are still, by and large, obscure. Although the behavioural/institutional stories that have been put forth are
persuasive and compatible with the bias observed in the holdings of mutual funds, there is no empirical smoking gun. We ourselves tend to believe in a 
universal ``lottery ticket'' \cite{Barberis} or embedded option mechanism that affects equally institutional investors (perhaps through the bonus optionality) and private investors, leading them 
to over-focus on potential spectacular upsides and forget much smaller but significant regular dividends. But this remains very much an open question.  

\vskip 0.5cm 

{\it{Acknowledgments}}: We thank F. Altarelli,  L. De Leo, A. Landier, D. Thesmar and P.-A. Reigneron for many insightful comments and for carefully reading the manuscript.

\section*{Appendix A: Pools of stocks}

The pools considered here are mostly the composition of standard indices for every given day, more precisely:

\begin{itemize}
\item Australia: 200 stocks of the S\&P/ASX 200
\item UK: 100 stocks of the FTSE 100
\item Europe: 600 most liquid (turnover in Euros) of the stocks mostly belonging to the SBF250 (France), CDAX (Germany), OMX (Sweden), SMI (Switzerland), IBEX (Spain), AEX (The Netherlands), FTSEMIB (Italy) indices, with a few stocks 
also coming from the Finnish, Norwegian, Belgian and Danish indices.
\item Japan:  500 most liquid stocks of the all-shares TOPIX index
\item Korea: 200 stocks of the KOSPI
\item Hong Kong: around 450 stocks of the Hang Seng Composite Index 
\item Canada: all stocks of the S\&P/TSX index
\item Brasil:  all stocks of the BOVESPA index (50 stocks in theory, but this number is actually variable in time -- 64 in 2015.)
\item S\&P500 and Russell 3000: stocks belonging to these two indexes.
\item US Small Caps: US stocks are ranked by their liquidity, and a pool is made with stocks of rank between 1501 and 2000.
\end{itemize}

Note that as a stock leaves the index, its position in the portfolio is liquidated at the next day's close price. 

\section*{Appendix B: Market neutrality vs. dollar neutrality}

The goal of this Appendix is to understand the origin of the bias in the low-vol strategies. The scalar product of the net $(1,1,....1)$ vector with the first
eigenvector times volatility is close to 95\% so there should be an indirect control of the Net Market Value (NMV) at least in principle. 
However, this is insufficient as the following toy-model reveals.

In a no costs Markowitz set-up we want to maximise
$$
\sum_i x_i p_i - \mu \sum_{i,j} x_i \sigma_i C_{ij} x_j \sigma_j
$$ where $x_i$
is the dollar end-of-day position, $p_i$ is a predictor, $\mu$ is the usual risk-aversion Lagrange parameter, $\sigma_i$ the volatility of
stock $i$, and $C_{ij}$ the normalized correlation matrix.
The Markowitz solution reads 
$$
x_i = \frac 1{2\mu\sigma_i} \sum_j C^{-1}_{ij} (p_j / \sigma_j) 
$$
We now chose a low-vol predictor as $p_i = a / \sigma_i$, where $a$
is a predictability scale. We are interested in computing NMV and Gross Market Value (GMV) in a simplified world where the correlation matrix has one (large) market
eigenvalue $\lambda^{(0)} \sim N \gg 1$ -- the rest of the spectrum being a Dirac mass around a small eigenvalue $\varepsilon^2$. Hence:
$$
C_{ij} = \sum_{\alpha} \lambda^{(\alpha)} v^{(\alpha)}_i v^{(\alpha)}_j \approx \lambda^{(0)} P^{(0)}_{ij} + \varepsilon^2
\sum_{\alpha=1}^{N-1} P^{(\alpha)}_{ij}
$$
where we have introduced the projectors on the eigenvectors $P^{(\alpha)}_{ij} = v^{(\alpha)}_i v^{(\alpha)}_j$. The inverse of this matrix is thus given by 
$$
C^{-1}_{ij} =
\frac 1 {\lambda^{(0)}} P^{(0)}_{ij} + \frac 1{\varepsilon^2} \sum_{\alpha=1}^{N-1} P^{(\alpha)}_{ij} 
$$
The trace condition on the correlation matrix is Tr $C$ = $N$ and implies that $1/\varepsilon^2 = (N-1) / (N-\lambda^{(0)})$. By using the complete spanning rule
$\delta_{ij} = \sum_{\alpha=0}^{N-1}P^{(\alpha)}_{ij}$ we end up with
$$
C^{-1}_{ij} \simeq \frac 1 {1 - \lambda^{(0)}/N} \left[ \delta_{ij} -
\left(1-\frac 1 {\lambda^{(0)}}\right) P^{(0)}_{ij} \right]
$$
Plugging this into the Markowitz solution yields to
$$
x_i = \frac a{2\mu}\frac 1 {1 -
\lambda^{(0)}/N} \left[ \sigma_i^{-3} - \left(1-\frac1{\lambda^{(0)}}\right) v^{(0)}_i\sigma_i^{-1}\sum_j v^{(0)}_j \sigma_j^{-2} \right] 
$$
The net market value is then NMV $= \sum_i x_i$ and depends on the prefactor $a/\mu$. This pre-factor will not matter in the ratio NMV
/ GMV but we can get rid of it anyway by using 
\begin{equation*} R^2 = \sum_{ij} x_i\sigma_i C_{ij} x_j \sigma_j \simeq \left(\frac
a{2\mu}\right)^2 \left(1-\frac {\lambda^{(0)}}{N} \right)^{-1} N \left( \langle \sigma^{-4}\rangle - \langle\sigma^{-2}\rangle^{2} \right)
\end{equation*} 
which leads to 
\begin{equation*} \text{NMV} = \frac{R}{\sqrt{(N - \lambda^{(0)})\left(\langle \sigma^{-4}\rangle -
\langle\sigma^{-2}\rangle^{2} \right)}} \left[ N \langle \sigma^{-3}\rangle - \left(1-\frac 1{\lambda^{(0)}} \right)\sum_i
v^{(0)}_i\sigma_i^{-1}\sum_j v^{(0)}_j \sigma_j^{-2} \right] 
\end{equation*}
We now use $\lambda^{(0)}\gg 1$ and assume $v^{(0)}_i \approx 1 / \sqrt{N}$. Then:
\begin{equation*} \text{NMV} \approx R N \sqrt{\frac{1}{(N - \lambda^{(0)})}} \,\, \frac{\langle y^3\rangle - \langle y\rangle \langle
y^2\rangle} {\sqrt{\langle y^4\rangle - \langle y^2\rangle^2}} \end{equation*} 
where we have introduced $y_i=1/\sigma_i$ and where we recall
that $\langle \cdots\rangle$ stands for a flat cross-sectional average over the stocks.
As a check, we can compute the risk exposure to the market mode: 
\begin{equation*} R^{(0)} = \sqrt{\lambda^{(0)}} \sum_i x_i\sigma_i v^{(0)}_i
\approx \frac{R}{\sqrt{\lambda^{(0)}(1 - \lambda^{(0)}/N))}} \,\, \frac{\langle y^2\rangle}{\sqrt{\langle y^4\rangle - \langle y^2\rangle^2}} 
\end{equation*}
which has the correct behaviour for large $\lambda^{(0)}$. The same kind of calculation can be done to get the $\text{GMV}=\sum_i |x_i|$
and one gets:
\begin{equation*}\text{NMV} \simeq RN \sqrt{\frac{1}{(N - \lambda^{(0)})}} \,\, \frac{\langle y |y^2- \langle y^2\rangle |\rangle}
{\sqrt{\langle y^4\rangle - \langle y^2\rangle^2}} 
\end{equation*}
So the net over gross ratio is 
\begin{equation*}
\frac{\textrm{NMV}}{\textrm{GMV}} = \frac{\langle y^3\rangle - \langle y\rangle \langle y^2\rangle}{\langle y |y^2- \langle y^2\rangle
|\rangle}, 
\end{equation*}
showing that this ratio is related to the higher moments of the cross-sectional volatility distribution, and is a priori non zero. 
One could assume the volatility distribution is an inverse Gamma distribution and compute the above number, or estimate directly these moments from the data. 
The average value lies in the 30-40\% range, and is thus much higher than naively thought. Hence, net dollar exposure is a real issue 
when dealing with low-vol portfolios.

\begin{figure}
\begin{center}
\epsfig{file=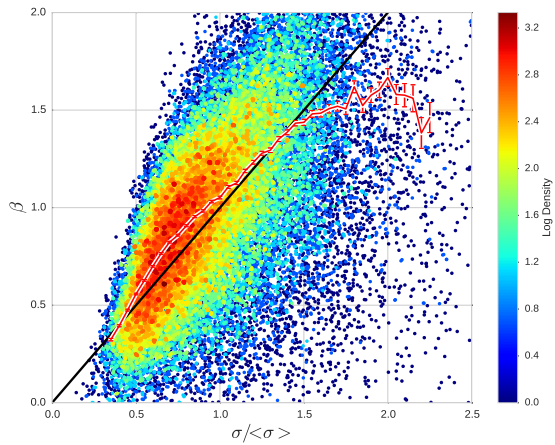, width=0.6\textwidth,angle=0}
\caption{\small{Comparison of the simple prediction, Eq. \ref{predictbeta}, with data using 2000 US stocks from 1971 to 2011. The black line corresponds to the model, while the 
purple line corresponds to a running average through the points. The color code indicates the density of data points.}}
\label{betaplot}
\end{center}
\end{figure}

\section*{Appendix C: Low-vol/low-$\beta$ correlation}

The correlation between low-vol and low-$\beta$ does not come as a complete surprise since both $\beta$ and volatility are supposed to be risk indicators 
pertaining to a given stock. 
On the other hand, the fact that using the market exposure of a stock gives almost exactly the same strategy as its volatility cries for an explanation. 
Here, we show that a simplified theoretical framework allows one to capture why this is the case.

We start from a standard one-factor model 
\begin{equation}
r_i = \beta_i \Phi + \varepsilon_i
\end{equation}
where the market factor $\Phi$ is defined for simplicity as the equi-weighted average stock return: $\Phi  = \sum_i r_i / N$, i.e. $\sum_i \beta_i = N$. 
We assume that $\Phi$ and $\varepsilon_i$ are uncorrelated (but not necessarily independent).
The exposure $\beta_i$ is computed from a linear regression and can be written as 
\begin{equation}
\beta_i  = \rho_{i,\Phi} \frac{\sigma_i}{\sigma_\Phi} \ ,
\end{equation}
with $\rho_{i,\Phi}$ the correlation between the market return and the return of stock $i$.

We now assume that in a first approximation all pair correlations are equal to a given $\rho_0$, i.e. the correlation matrix has the form 
$\rho_{i,j} = \delta_{i,j} + \rho_0 (1-\delta_{i,j})$. This of course neglects sector effects but is not completely off the mark. 
Up to $1/N$ corrections, the market volatility is then simply given by $\sigma^2_\Phi = \rho_0  \sigma_{\text{av.}}^2$, where 
$ \sigma_{\text{av.}} = \sum_i \sigma_i / N$ is the cross-sectional volatility average. Similarly, the stock/market correlation can be computed, up to $1/N$ corrections, as:
\begin{equation}
\rho_{i,\Phi} = \frac{\langle r_i \frac 1N \sum_j r_j \rangle }{\sigma_i \sigma_\Phi} \approx \frac{\rho_0  \sigma_i \sigma_{\text{av.}}}{\sigma_i \sqrt{\rho_0} \sigma_{\text{av.}}} 
= \sqrt{\rho_0} \ .
\end{equation}
Hence within this highly schematic framework we find the simple relation:
\begin{equation} \label{predictbeta}
\beta_i = \sqrt{\rho_0} \frac{\sigma_i}{\sigma_\Phi} \equiv \frac{\sigma_i}{\sigma_{\text{av.}}} \ ,
\end{equation}
which could have been guessed by symmetry.

This result is interesting for several reasons: (i) it relies on the simplified hypothesis on pair-correlations but it does not require this average correlation $\rho_0$ to be known, 
nor to be constant over time; (ii) it suggests a natural scaling that one should use when looking at large, heterogeneous data sets and, finally,
(iii) it shows that $\beta$ and $\sigma$ are simply proportional one to each other, and therefore that low-vol and low-$\beta$ are bound to be highly correlated. 
As shown in Fig. \ref{betaplot}, our simple prediction is indeed backed by data: we compare the model to the data obtained for 2000 US stocks from 1971 to 2011, where each point 
corresponds to a pair $\sigma_i,\beta_i$. The regime where $\beta$ saturates for $\sigma$ large is not explained by the model and it would require more sophisticated assumptions. 
As shown by the density of points, though, this regime only represents a small part of the data set.

Note that within the same framework, one finds that the idiosyncratic volatility $\langle \varepsilon_i^2 \rangle$ is given by:
\be
\langle \varepsilon_i^2 \rangle = \sigma_i^2 - \beta_i^2 \sigma_\Phi^2 = \beta_i^2 \left[ 1 - \rho_0 \right]  \sigma_{\text{av.}}^2,
\ee
showing that the idiosyncratic volatility is also strongly correlated to the $\beta$.

\end{document}